\documentclass[aps,twocolumn,prd,showpacs,floatfix]{revtex4}
\usepackage{epsfig}
\usepackage{bm}
\usepackage{dcolumn}
\usepackage{multirow}

\def\reff@jnl#1{{\rm#1\/}}

\def\aj{\reff@jnl{AJ}}                  % Astronomical Journal
\def\araa{\reff@jnl{ARA\&A}}            % Annual Review of Astron and Astrophys
\def\apj{\reff@jnl{ApJ}}                        % Astrophysical Journal
\def\apjl{\reff@jnl{ApJ}}               % Astrophysical Journal, Letters
\def\apjs{\reff@jnl{ApJS}}              % Astrophysical Journal, Supplement
\def\ao{\reff@jnl{Appl.Optics}}         % Applied Optics
\def\apss{\reff@jnl{Ap\&SS}}            % Astrophysics and Space Science
\def\aap{\reff@jnl{A\&A}}               % Astronomy and Astrophysics
\def\aapr{\reff@jnl{A\&A~Rev.}}         % Astronomy and Astrophysics Reviews
\def\aaps{\reff@jnl{A\&AS}}             % Astronomy and Astrophysics, Supplement
\def\azh{\reff@jnl{AZh}}                        % Astronomicheskii Zhurnal
\def\baas{\reff@jnl{BAAS}}              % Bulletin of the AAS
\def\jrasc{\reff@jnl{JRASC}}            % Journal of the RAS of Canada
\def\memras{\reff@jnl{MmRAS}}           % Memoirs of the RAS
\def\mnras{\reff@jnl{MNRAS}}            % Monthly Notices of the RAS
\def\pra{\reff@jnl{Phys.Rev.A}}         % Physical Review A: General Physics
\def\prb{\reff@jnl{Phys.Rev.B}}         % Physical Review B: Solid State
\def\prc{\reff@jnl{Phys.Rev.C}}         % Physical Review C
\def\prd{\reff@jnl{Phys.Rev.D}}         % Physical Review D
\def\prl{\reff@jnl{Phys.Rev.Lett}}      % Physical Review Letters
\def\pasp{\reff@jnl{PASP}}              % Publications of the ASP
\def\pasj{\reff@jnl{PASJ}}              % Publications of the ASJ
\def\qjras{\reff@jnl{QJRAS}}            % Quarterly Journal of the RAS
\def\skytel{\reff@jnl{S\&T}}            % Sky and Telescope
\def\solphys{\reff@jnl{Solar~Phys.}}    % Solar Physics
\def\sovast{\reff@jnl{Soviet~Ast.}}     % Soviet Astronomy
\def\ssr{\reff@jnl{Space~Sci.Rev.}}     % Space Science Reviews
\def\zap{\reff@jnl{ZAp}}                        % Zeitschrift fuer Astrophysik
\def\nat{\reff@jnl{Nature}}             % Nature 
\newcommand{\CC}{\mathsf{C}}
\newcommand{\DD}{\mathsf{D}}

\begin{document}

\title{Assessing the effects of foregrounds and sky removal in WMAP} 
\author{An\v{z}e Slosar} 
\affiliation{Faculty of Mathematics and Physics, University of Ljubljana, Slovenia}
\author{Uro\v{s} Seljak}
\affiliation{Department of Physics, Princeton University, Princeton NJ 08544, U.S.A.}

\date{\today}

\begin{abstract}

Many recent analyses have indicated that large scale WMAP data display
anomalies that appear inconsistent with the standard cosmological
paradigm.  However, the effects of foreground contamination, which
require elimination of some fraction of the data, have not been
fully investigated due to the complexity in the analysis.  Here we
develop a general formalism of how to incorporate these effects in any
analysis of this type.  Our approach is to compute the full
multi-dimensional probability distribution function of all possible
sky realizations that are consistent with the data and with the
allowed level of contamination. Any statistic can be integrated over
this probability distribution to assess its significance.  As an
example we apply this method to compute the joint probability
distribution function for the possible realizations of quadrupole and
octopole using the WMAP data. This 12 dimensional distribution
function is explored using the Markov Chain Monte Carlo technique. The
resulting chains are used to asses the statistical significance of the
low quadrupole using frequentist methods, which we find to be 3-4\%.
Octopole is normal and the probability of it being anomalously low 
or as low as WMAP reported value is very small. 
We address the quadrupole-octopole alignment using several methods
that have been recently used to argue for anomalies, such as angular
momentum dispersion, multipole vectors and a new method based on
feature matching. While we confirm that the full sky map ILC suggest an
alignment, we find that removing the most contaminated part of the
data also removes any evidence of alignment: the probability
distributions strongly disfavor the alignment. This suggests that most
of the evidence for it comes from non-Gaussian features in the part of
the data most contaminated by the foregrounds.  We also present an 
example, that of octopole alignment with the ecliptic, 
where the statistical significance can be enhanced by removing the 
contamination. 

\end{abstract}


\topmargin-1cm

\maketitle

\section{Introduction}

The large scale structure of the WMAP data \citep{2003MNRAS.346L..26E}
has received a lot of attention since the first data release. Some
authors have focused on the seemingly low values of quadrupole and
octopole
\cite{2003MNRAS.346L..26E,2004MNRAS.348..885E,2003PhRvD..68l3523T,2004astro.ph..3073S},
and their alignment \citep{2004PhRvD..69d063516,2004astro.ph..3353S},
while others considered various asymmetries in the data
\citep{2003astro.ph..7507E,2004astro.ph..3098E,2004astro.ph..2396H,2004astro.ph..4206H}.
Some of these analyses are performed on one of the available full sky
WMAP maps, either the original WMAP Internal Linear Combination Map
(ILC) or an alternative map \cite{2003PhRvD..68l3523T}, which we will
refer to as TOH map. The full sky maps have the advantage that the harmonic
analysis is unique, which facilitates investigation and assessment of
statistical significance. We note that \cite{2004astro.ph..3098E}
prudently warn against the usage of the foreground corrected full-sky
maps and use appropriate Monte-Carlo simulations.

However, full sky maps are not free of contamination, as was clearly
emphasized by the WMAP team warning that their ILC map should not be
used for science purposes. These are dominated by galactic foregrounds
such as dust, synchrotron or free-free emission.  For this reason the
power spectrum analysis is done on cut sky, where about 15-25\% of
most contaminated data in the galactic plane is removed.  Even outside
this region there are residual uncertainties associated with
imperfections of the foreground removal.  If ignored they may cause
spurious alignments or other anomalies that appear statistically
significant under the assumption that CMB is a Gaussian random field.

In this paper we revisit the statistical significance of these tests
using a different approach.  Rather than ignoring the effects of
foregrounds we try to take them into account explicitly by
exploring the uncertainties they induce in the measurements of the
multipole moments $a_{\ell m}$.  We assess this uncertainty by
determining the joint multi-dimensional probability distribution
function for the true sky multipole moments. Once these are determined
we can apply them to the statistics of choice to obtain their values
in the presence of uncertainties associated with imperfect foreground
removal or sky cuts.  We do not address the question of the meaning of
a given statistic: all of the statistics are a-posteriori and their
statistical significance is difficult to assess. Instead, our goal is to compare
the values of these statistics with and without the inclusion of
foreground uncertainties to see if including the latter changes the
conclusions significantly.  

In this paper we are interested in large scale features, so we focus
exclusively on quadrupole and octopole.  This allows us to perform
several tests.  First, we can revisit the question of whether the
quadrupole and octopole are low in the presence of additional
uncertainties associated with the foregrounds and sky cuts.  Secondly,
we can test how robust results of methods for measuring the alignment
of quadrupole and octopole are, once these uncertainties are taken
into account.  We do this by applying several of existing statistics,
as well as a new one we developed. Finally, we also explore the
alignment of the large scale features with specific directions in the
sky, such as that of ecliptic plane. 

\section{Method}

We need to distinguish between two types of uncertainties. 
The theory of CMB fluctuations can only predict ensemble averages of
the power spectrum of CMB fluctuations. Given the statistical
description of initial fluctuations in the form of a power spectrum
and parameters of a given cosmological model it is only possible to
predict the power spectrum of the CMB fluctuations averaged over all
possible realizations of the universe. Since there is only one CMB sky
that we can directly measure, there is a corresponding uncertainty in
our inference of the true cosmic power spectrum. This uncertainty is
often referred to as cosmic variance.

In terms of multipole moments $a_{\ell m}$, for an idealized
experiment with full sky coverage and no foreground contamination or
noise, their values can be determined precisely. In this case the only
uncertainty is the cosmic variance.  However, if there is noise or
contamination we may not be able to measure precisely the individual
multipole moments either. This introduces an additional level of
uncertainty.  For the power spectrum analysis such an uncertainty is
automatically accounted for in the likelihood analysis, where sky-cuts
and foreground uncertainties can be included in the likelihood
calculations without determining the actual values of multipole
moments. For other more complicated statistics one must determine the
probability distribution of the multipole moments first and then 
apply these to the statistic of interest.

\subsection{Probability distribution of the multipole moments}

In general determining the full probability distribution of multipole
moments can be numerically challenging, as the multipole moments will
be correlated and their number will grow as the square of the number
of considered multipoles. Here we focus on large scales only, where
quadrupole and octopole dominate.  Their joint probability
distribution function has $12$ degrees of freedom.  
We use Markov Chain Monte Carlo
in the form of the simplest Metropolis algorithm
\citep{Metropolis53equations} with a fixed Gaussian width proposal
function. Since the likelihood evaluations are very fast this simple
method is sufficient.
We note that the posterior
distribution $p(a_{\ell m})$ is gaussian by construction and 
can be fully constrained with just a few likelihood
evaluations. This can considerably speed up the calculations, but 
is not really needed for our analysis where we focus on the lowest 
multipoles only. 

In order to make valid constraints on various realizations of the
quadrupole and octopole we need to devise a way of calculating the
likelihood of a given $a_{2m}$s and $a_{3m}$s in the presence of the
instrumental noise, fluctuations due to the power in the higher
multipoles and foreground contamination. In analogy with the common
$\chi^2$ analysis, the likelihood of a given model is the probability
that the residuals between the theoretical map for a given theoretical
model (i.e. a map corresponding to a quadrupole and octopole) and the
measured map are a possible noise realization. Therefore, the likelihood can
be written as:
\begin{equation}
  \log L = -\frac{1}{2} \mathbf{(d-d_T)^T (C^{\rm total})^{-1}
    (d-d_T)} + D
%%%- \frac{1}{2}\left(\log |\mathbf{C}| + N \log 2\pi\right),
\label{eq:2}
\end{equation}
where $\mathbf{d_T}$ is the theoretical map data-vector
\begin{equation}
  (\mathbf{d_T})_i = \sum_{\ell=2,3} \sum_{m=-\ell \ldots \ell} Y_{\ell m}
  (\mathbf{n}_i) a_{lm},
\end{equation}
with $\mathbf{n}_i$ being the unit vector in the direction of the $i$-th data-point
and $D$ is an unimportant constant.  

The covariance matrix
$\mathbf{C^{\rm total}}$ is the total covariance matrix, which can be
broken into several parts as follows:
\begin{eqnarray}
\label{eq:lambda}
  \mathbf{C}^{\rm total} = \mathbf{C} + \mathbf{N} + \lambda (
\mathbf{C}^{\rm dust}+\mathbf{C}^{\rm synch}+\mathbf{C}^{\rm
free-free}&+& \nonumber \\ \mathbf{C}^{\ell=0}+\mathbf{C}^{\ell=1}
)&,&
\end{eqnarray}

The main contribution is the noise due to fluctuations in the
multipoles higher than recovered ones:
\begin{equation}
  C_{i,j} = \sum_{\ell=4}^{\infty} \frac{2\ell+1}{4\pi}  C_\ell
  P_\ell(\cos \theta_{i,j})B_\ell, 
\label{eq:1}
\end{equation}
where the sum starts at the lowest multipole which is not being
recovered, $P_\ell$ is the Legendre Polynomial of order $\ell$,
$B_\ell$ is the beam smoothing and $\theta_{i,j}$ is the angle between
$i$th and $j$th point on the sky. We denote with $\mathbf{N}$ the instrumental
noise matrix, which for $l=2,3$ is small compared to other sources of noise. 

The matrices $\mathbf{C}^{\rm dust}$, $\mathbf{C}^{\rm synch}$ and
$\mathbf{C}^{\rm free-free}$ are foreground contamination matrices,
given by
\begin{equation}
  \mathbf{C}^{\rm foreground} = \mathbf{LL}^{\dagger},
\end{equation}
where $\mathbf{L}$ is the foreground template vector. As discussed in
previous paper \cite{2004astro.ph..3073S}, the linear modes that
correlate with the foreground template vector are effectively
marginalized out in the limit of $\lambda \rightarrow \infty$. This is
not the only option if one has additional information on the probable
amplitude of these foregrounds, so in this paper we also try
$\lambda=1$, corresponding to the case where multi-frequency 
information in WMAP does provide some information on their amplitude, but 
with an error of order unity compared to the best fit value. 
Finally, we also marginalize the residual dipole and monopole
in the map with $\lambda \rightarrow \infty$, since we have 
no external information on their value.

This method allows us to calculate the likelihood of a given
realization of a quadrupole and octopole in the presence of power in
higher multipoles, foregrounds and monopole/dipole contamination.  We
explore the likelihood surface by making steps in a random direction,
at each step deciding on whether to accept the new event based on the
likelihood ratio relative to previous event. This MCMC sampling of the
likelihood surface results in a 12 dimensional probability
distribution of multipole moments, which is consistent with the data
and with the allowed level of foreground contamination.  Each MCMC
element consists of a realization of multipole moments defined on the
{\it uncontaminated and uncut sky}, so one can apply any statistic of
choice to a given realization without having to worry about
contamination, noise or sky cuts and various effects associated with
implied priors.  By averaging over MCMC elements one is performing
marginalization over the full probability distribution.
A related method has been proposed in
\cite{2003astro.ph.10080W} and \cite{2004astro.ph..1623W}.

It is important to realize that the inverse of the covariance matrix
is calculated just once and therefore the likelihood evaluation is a
very fast process. In fact, the entire MCMC process takes a few hours
on a modern PC workstation. The method is trivially extended to higher
multipoles. If a map has $N$ pixels, it is completely defined by the
$\ell_{\rm max} \propto N^{1/2}$ multipoles. Assuming the time
required to calculate a candidate map scales as $N$ (if we are moving
in a specific $a_{\ell,m}$ direction) and that the MCMC chain is limited
by the random walk rather than shot noise in the sample density, then
a very favorable scaling of $N^{3/2}$ is obtained for reconstructing
all multipoles in the map. However, it is very likely that the method
will be limited by the very complicated multi-modal likelihood shape
in the case of higher multipole reconstruction.

\subsection{Choice of maps and foreground templates}

We focus on three increasingly conservative combinations to asses
the stability of statistical inferences. All maps are
smoothed using $5^\circ$ FWHM beam and down-sampled to the
\texttt{nside=16} Healpix map.
Our three datasets in order of increasing conservativeness are:

\begin{itemize}
\item \texttt{ILC} dataset contains the full-sky ILC map. Noise is
  ignored and so are foreground contaminations.  This dataset can be
  used to compare our results with the previously reported results and
  to asses whether the methods produces the expected results, but is 
likely to be too contaminated for results to be reliable.

\item \texttt{Wd} dataset takes the W channel data, applies the Kp2
  mask, subtracts the  MEM derived  Free-Free foreground and
  marginalizes over MEM derived Dust template (using 
$\lambda=1$ and $\lambda=\infty$).  

\item \texttt{Vdfs} takes the V channel map, applies the Kp2 mask,
  and marginalizes over all three foreground templates of Dust, Free-Free
  and Synchrotron emission.
\end{itemize}

We have run the MCMC process until at least 500000 independent samples were
obtained and discarded first 10000 samples. The process is very fast
and a large number of samples ensures that the chain is well
converged. We checked this by comparing results of the first quarter of samples
with that of the last quarter.

\subsection{Priors and implied priors}
\label{sec:impl}

 Multipole moments $a_{\ell m}$ are drawn from a Gaussian distribution
with variance $C_\ell$. The values of $a_{\ell m}$ are in general
complex, but the requirement that the observed sky is a real quantity
demands that $a_{\ell m}=a_{\ell -m}^{\ast}$. This requires that the
imaginary part of $a_{\ell 0}$ vanishes and thus we are left with
$2\ell + 1$ degrees of freedom.  We introduce the symbol $D_\ell$
to describe the ``measured power spectrum'' on a given sky,
\begin{equation}
  D_\ell = {1 \over 2l+1} \sum_m  |a_{\ell m}|^2 .
\label{dl}
\end{equation}
Our MCMC process takes flat priors on the values of each of the
imaginary and real components of $a_{\ell m}$, which
are wide enough so that they do not affect the posterior probability
distribution $p(a_{\ell m})$. This, however, implies a non-flat prior on
the derived probability distribution for $D_\ell$. 

This can be calculated as follows. In the $2\ell+1$ dimensional space
spanned by $(a_{\ell,0}$, $\sqrt{2} {\rm\ Re\ } (a_{\ell,1})$,
$\sqrt{2} {\rm\ Im\ } (a_{\ell,1})$, \ldots $)$ the points of constant
$D_\ell$ lie on the hyper-sphere with radius $R= ((2\ell+1)
D)^{1/2}$. The number of states corresponding to the volume of the
shell of thickness ${\rm d}R$ determines the number of available
states and therefore the implied prior is given by

\begin{equation}
  p_{\rm implied}(D_\ell){\rm d}D_{\ell}\propto R^{2\ell}{\rm
  d}R\propto D^{\ell-\frac{1}{2}}{\rm d}D.
\end{equation}

In particular, the implied prior for the quadrupole is
$p_{\rm implied}(D_2) \propto D^{3/2}$ and for the octopole $p_{\rm
implied}(D_3) \propto D^{5/2}$.

As usual in such analyses, if the data were perfect the assumed prior 
would be irrelevant, while in the opposite limit all information comes 
from priors. It seems however unlikely that the assumed priors would 
affect our results on alignments (discussed below), as long as 
the priors do not introduce any correlations between the multipole moments. 
Therefore, we take these implied priors into account when calculating the
estimates of the statistical significance of quadrupole and octopole
(because a flat $C_\ell$ prior is usually assumed in analyses of this
kind), but not when assessing the alignment of quadrupole and octopole.

In the rest of this paper we will also use symbols $\CC_\ell$ and
$\DD_\ell$ which correspond to a more conventional normalization of
$C_\ell$ and $D_\ell$
\begin{eqnarray}
  \CC_\ell = \frac{C_\ell \ell (\ell+1)}{2\pi} \\
  \DD_\ell = \frac{D_\ell \ell (\ell+1)}{2\pi}. 
\end{eqnarray}

\subsection{Consistency check: measured versus true power spectrum}

The $D_\ell$ (equation \ref{dl}) is $\chi^2$ distributed with $2\ell+1$ degrees of freedom,
\begin{equation}
  p(D_\ell| C_\ell) = \left[ C_\ell \Gamma
  \left(\ell+\frac{1}{2}\right) \right]^{-1} \exp\left(\frac{D_\ell}{C_\ell}\right)
  \left[\frac{D_\ell}{C_\ell}\right]^{\ell-\frac{1}{2}}.
\label{eq:aa}
\end{equation}

However, we measure $D_\ell$ and assuming a flat prior on $C_\ell$ the
Bayes theorem says:
\begin{equation}
  p(C_\ell|D_\ell) \propto p(D_\ell | C_\ell)
\label{eq:bb}
\end{equation}

For an idealized experiment with full sky coverage and no foreground
contamination or noise, the value of $D_\ell$ can be determined
precisely and therefore one is only constrained by the cosmic variance
from Equations (\ref{eq:aa}) and (\ref{eq:bb}) as discussed in the
introduction.
 For a real experiment, one must marginalize over this uncertainty
\begin{equation}
  p(C_\ell) = \int p(C_\ell|D_\ell) p(D_\ell) {\rm d}p(D_\ell)
\label{eq:cc}
\end{equation}

Given the probability distribution function for 
$a_{\ell m}$ allows one to calculate the probability
distribution function for $p(D_\ell)$ using
\begin{eqnarray}
  p(D_\ell) = \int \cdots \int \frac{\sum_{m} |a_{\ell m}|^2}{2 \ell
+1 } P\left(a_{00},\ldots a_{\ell_{\rm max} \ell_{\rm max}}\right)
\nonumber \\ \times {\rm d}a_{\ell 0}\ldots {\rm d}a_{\ell \ell},
\end{eqnarray}
where $\ell_{\rm max}$ is the maximum $\ell$ for which this distribution is
recovered. Combining the expressions above gives us $p(C_\ell)$. 

\begin{figure}
\epsfig{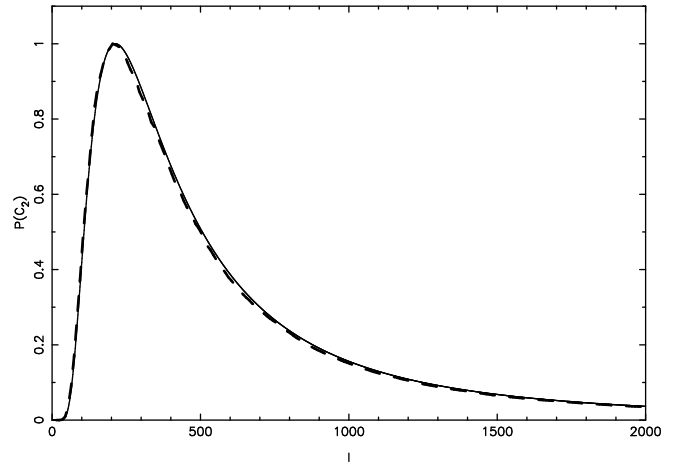}
\caption{\label{fig:ih}This Figure shows the probability distribution
  function for the true quadrupole in Vdfs case, calculated using two
  methods. The solid line correspond the $p(\CC_2)$ distribution
  function derived using MCMC chains, while the dashed line corresponds to the
  distribution derived using the exact likelihood calculation by
  matrix inversion. The two curves are in a good agreement indicating that the
  method is numerically robust.}
\end{figure}

As mentioned in the introduction the various power spectrum
determination procedures such as PCL (see
e.g. \citep{2002ApJ...567....2H}) and QML (see
e.g. \citep{1997PhRvD..55.5895T}) estimators or the exact methods
using matrix inversion already take into account the cosmic variance
to derive $p(C_\ell)$.  As a consistency check we compare the
$p(\CC_2)$ derived from our chains to that of the exact likelihood
analysis performed in \cite{2004astro.ph..3073S}.  We do this in the
following manner. We calculate the $p(\DD_2)$ using our MCMC
chains. The inferred probability distribution is then multiplied by
$D^{-3/2}$ to take into account the effect of the implied prior effect
as described in the section \ref{sec:impl}. Finally we numerically
integrate Equation (\ref{eq:cc}) to get the probability distribution
$p(\CC_2)$.  The resulting $p(\CC_2)$ for the Vdfs case is shown in
Figure \ref{fig:ih}, together with the $p(\CC_2)$ obtained by
calculating the exact likelihood using the matrix formalism. The two
curves are in a good agreement. This confirms that the method works as
expected.

\subsection{Fitting individual foregrounds}
\label{sec:fitt-indiv-foregr}
\begin{figure}
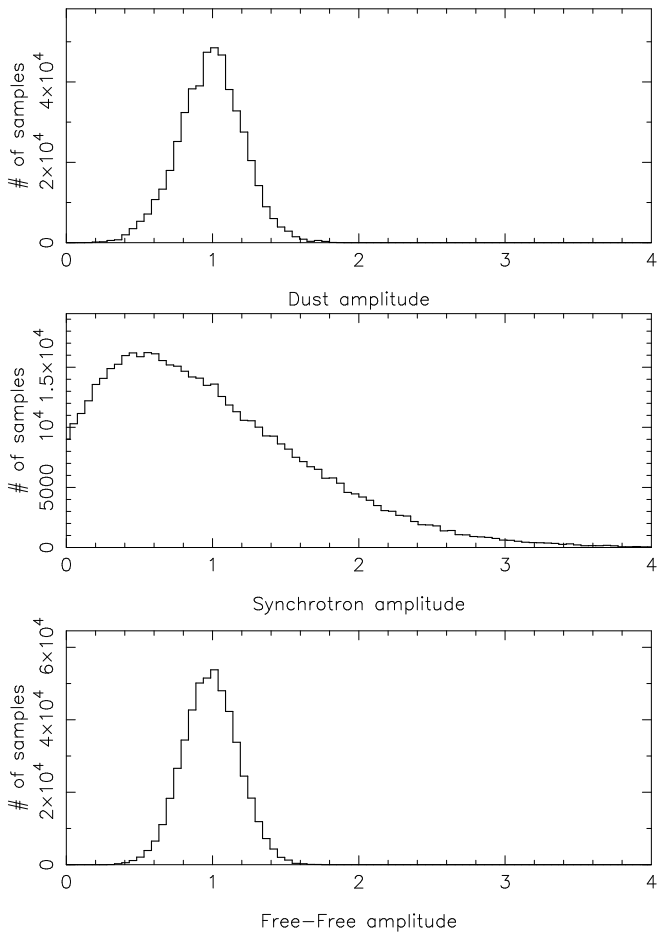

  \epsfig{file=dust-temp.eps, height=\linewidth, angle=-90}
  \epsfig{file=synch-temp.eps, height=\linewidth, angle=-90}
  \epsfig{file=ff-temp.eps, height=\linewidth, angle=-90}

\caption{\label{fig:amplfit} This figure shows the inferred amplitudes
of foreground maps when used to fit the data. The x axis has been
rescaled to average to one. See text for further discussion. }
\end{figure}

We have also tested the behavior of the system if one takes the
amplitudes of the foregrounds to be additional variables of the
system. In this case, the marginalization is performed by explicit fit
to the amplitudes of these foregrounds. We have taken the V frequency
data, applied the KP2 cut and used the standard foreground templates
used by the WMAP team: the 408MHz Haslam synchrotron radiation map
\citep{1982A&AS...47....1H},H-$\alpha$ map from
\cite{2003ApJS..146..407F} as a tracer of free-free emission and the
FDS dust template based on \cite{1999ApJ...524..867F}. We used these
templates instead of MEM derived maps to make the inferred amplitudes
as statistically independent as possible and thus avoid complications
associated with signal-noise correlations which become apparent when using MEM
foreground maps. Additionally we imposed constraints that all
amplitudes must be positive. The resulting 12+3 dimensional
probability space is explored using the standard MCMC procedure. We
show the resultant amplitudes in Figure \ref{fig:amplfit}. In this
plot, the amplitudes have been rescaled to average to one as the
absolute numbers are not important for our application (templates give only the flux ratio
between the pixels).  The data determine the amplitude of the dust and
free-free emissions fairly accurately, but the amplitude of the
synchrotron is only poorly constrained and consistent with 0. This is
what is expected, since the synchrotron is not anticipated to be a major
contaminant at the frequencies of the V channel.

\begin{figure}
  \epsfig{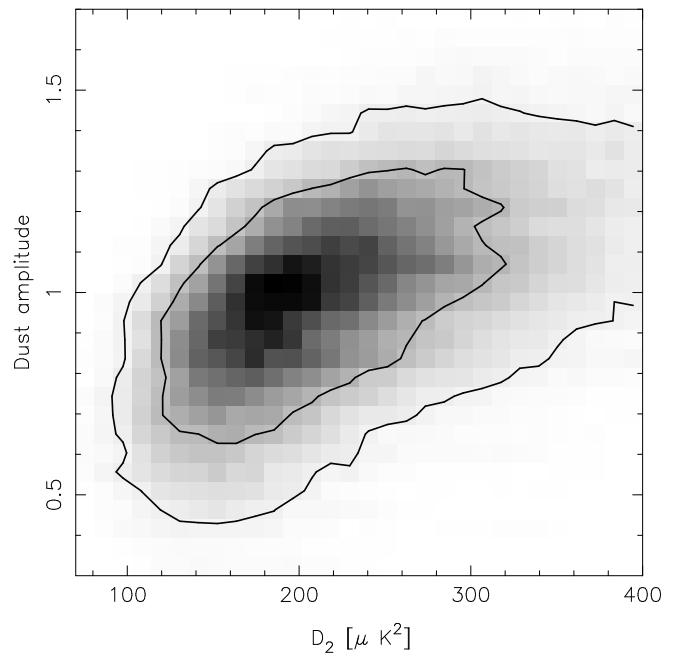}

\caption{\label{fig:ampl2d} This Figure shows the two-dimensional
  distribution of the probability on the  $\DD_2$ - Dust amplitude
  plane. Contours correspond to 1 and 2-sigma confidence limits. See
  text for further discussion.
   }
\end{figure}

We have also investigated degeneracy directions which might exist
between various templated amplitudes and the values of $\DD_2$ and
$\DD_3$. Template amplitudes are nearly completely uncorrelated between each
other and with the multipole amplitudes. The only marginally
significant correlation exists between the Dust amplitude and the
value of $\DD_2$. This is shown in Figure \ref{fig:ampl2d}.

\section{Results}

\subsection{Distributions for $\DD_2$ and $\DD_3$}

\begin{figure}
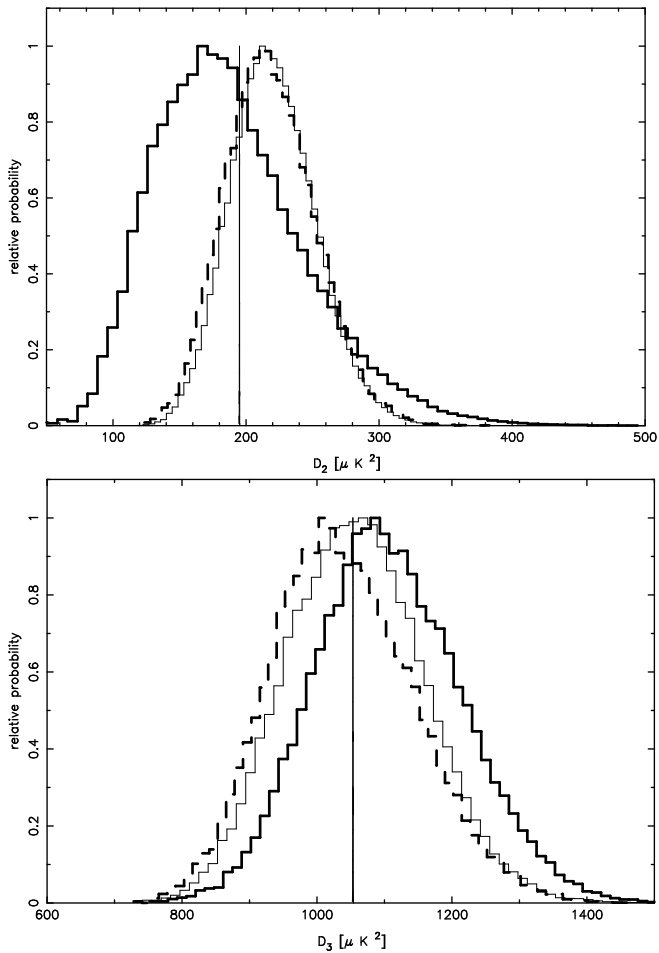

  \epsfig{file=p1.eps, height=\linewidth, angle=-90}

  \epsfig{file=p2.eps, height=\linewidth, angle=-90}

\caption{\label{fig:p12} This Figure shows the distribution for
  $\DD_2$ and $\DD_3$ for the following four cases: ILC (vertical lines), Wd
  (thin line), Wd with $\lambda=1$ (thick dashed line) and Vdfs (thick line).}
\end{figure}

In Figure \ref{fig:p12} we show the marginalized one-dimensional
distributions for $\DD_2$ and $\DD_3$ for the three cases and the
\texttt{Wd} case with $\lambda=1$. These show several interesting
features. The ILC map gives values of $\sim 195 \mu K^2$
for the quadrupole and $\sim 1050 \mu K^2$ for the octopole, in a good
agreement with results obtained previously on the same map using the
QML estimator \cite{2004MNRAS.348..885E}.  Zooming up into the region
actually reveals the probability distribution is a narrow Gaussian
with FWHM of a few $\mu K^2$, consistent with the effect of finite
pixelization, etc. The confidence limits on the two parameters widen
with the inclusion of sky cuts and the increasing amount of foreground
correlated modes being marginalised out. We note that \texttt{Wd} case
is very similar for $\lambda=1$ and $\lambda\rightarrow \infty$. This
is consistent with our findings in Section
\ref{sec:fitt-indiv-foregr}. The former sets the uncertainty in the
foreground amplitude to the $O(1)$, while the latter effectively
marginalises over the foreground amplitude probability given the data,
which we have shown to be of the order $O(1)$. Therefore, we will
limit our discussion to $\lambda \rightarrow \infty$ in the rest of
this paper. It is interesting to note that while including more
uncertainty in foreground subtraction (Vdfs versus Wd) makes the
distribution broader, it also pushes the median to lower values.

\begin{figure}
  \epsfig{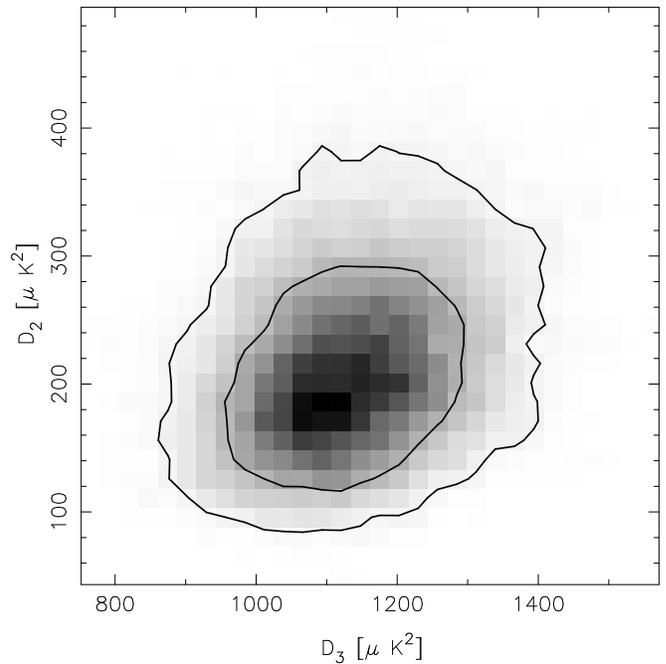}

\caption{\label{fig:DDD} This Figure shows the two-dimensional
  distribution of the probability on the  $\DD_2$-$\DD_3$ plane for
  the Vdfs case. Contours correspond to one-sigma and two sigma
  confidence limits. }
\end{figure}

We also plot the 2-dimensional distribution on the $\DD_2$-$\DD_3$
plane for the most conservative Vdfs case in Figure
\ref{fig:DDD}. $\DD_2$ and $\DD_3$ are nearly completely uncorrelated,
in accordance with previous findings.

\subsection{How low are the quadrupole and octopole?} 

Reference \cite{2003MNRAS.346L..26E} proposes two methods for
assessing the statistical significance for the low quadrupole,
assuming that the $\DD_2$ distribution is known.  The Bayesian
estimate is concerned with the probability that $\CC_2$ is greater
that its concordant value ($1150 \mu K^2$) given the value of
$\DD_2$. This is just the integral of the $p(\CC_2)$ from its
concordant value upwards and therefore our MCMC chains do not add
new information, as the $p(\CC_2)$ can already be obtained from
the exact likelihood evaluation shown in Figure 5 of the reference
\cite{2004astro.ph..3073S}. Here we highlight the fact that the
likelihood at high values of true quadrupole is very slowly decreasing
and as a result the probability depends on the adopted prior.  We
quote these results in the Table 1.

\begin{table*}
\noindent
{\footnotesize Table 1: This table shows the Bayesian estimates for
  the probability of the low quadrupole assuming a narrow (uniform
  between 0 and 2000 $\mu K^2$) and a wide (uniform between 0 and 10000 $\mu
  K^2$) prior on the value of $\DD_2$. }
\begin{center}
\begin{tabular}{ccc}
\hline
    Dataset & Narrow prior &  Wide prior\\
\hline 
    $p(\DD_2)=\delta(\DD_2-123 \mu K^2)$ & 5.0\%  & 8.5\% \\
 ILC   & 9.1\% & 16\%\\
 Wd & 11\% & 18\%\\
 Vdfs  & 8.8\% & 15\%\\

\hline
  \end{tabular}
\end{center}

\end{table*}

Frequentist estimate is concerned with the probability that the
value of $\DD_2$ is as small or smaller than the observed $\DD_2$  given that the
value of $\CC_2$ takes its concordant value ($1150 \mu K^2$). We
calculate a ``frequentist'' estimate integrating over the probability
distribution for $p(\DD_2)$. We emphasize that this is not a real 
frequentist estimate, since a true frequentist statistic is never 
a-posteriori. We have also corrected for the Bayesian implied prior on
the values of $\DD_2$ as discussed in the Section \ref{sec:impl}. 

% (this concern is much more of an issue for other statistics
%described below). 
%In addition, we have assumed flat priors on multipole 
%moments $a_{lm}$'s, so we still have a Bayesian component in our
%analysis.

The results are shown in Table 2. We note that the values are a
factor of a few higher than the original WMAP estimate and do not
depend much on which method we use.  The ILC value of $\DD_2\sim 195
\mu K^2$ is somewhat higher, but consistent with the original WMAP value of $154
\pm 70 \mu K^2$ \cite{2003ApJS..148....1B}. We note, however, that the
number quoted in the official power spectrum file is considerably
lower, $123 \mu K^2$. 
The probability for the frequentist estimate rises from 1\%
to about 3\% with the larger value, 
which is already a major increase. Taking into
account the non-negligible width in the probability distribution
function for $\DD_2$ for the Wd and Vdfs does not change the
probability significantly. Note that the WMAP value of $123 \mu K^2$
is perfectly acceptable for the more conservative Vdfs analysis, which
marginalizes over 3 foreground templates, but appears nearly excluded
from Wd analysis.  In other words, if one assumes W map with KP2
mask and assumes the foreground uncertainties are associated only with
the dust template then the probability for the actual quadrupole on
the sky $\DD_2$ to be below $123\mu K^2$ is extremely small, less than
0.01\%.

Octopole results show a similar pattern, except that octopole is 
not low compared to standard model. The ILC value of $\sim 1050
\mu K^2$ is in near perfect agreement with the concordance model value 
of around 1100$\mu K^2$. 
This is changed into a broader distribution by the more
conservative treatments, with the peak values close to the ILC value
around 1100$\mu K^2$ and the width of about 150$\mu K^2$ FWHM. Both
distributions are similar, so the details of foreground
marginalization do not appear very important here.  However, the low
octopole value reported by WMAP, $611 \mu K^2$, is not within the
allowed range even for the most conservative treatment and there is
not a single MCMC sample that would give a value this low.  It is not
clear what the cause for this discrepancy with WMAP values is, but one
possible culprit is the estimator used by WMAP, which is noisier than
the optimal maximum likelihood estimator and which thus can lead to an
estimate that differs significantly from the actual value \cite{2004MNRAS.348..885E}.
In conclusion, quadrupole is somewhat but not anomalously 
low, its probability is around 3-4\%,
while octopole is perfectly normal and there is essentially zero probability 
for it to be as low as WMAP reported value. 

\begin{table}
\noindent
{\footnotesize Table 2:   This table shows the ``frequentist'' estimates of
  the probability of low quadrupole. See text for further discussion.}
\begin{center}

  \begin{tabular}{cc}
\hline
    Dataset & Frequentist probability \\
\hline
    $p(\DD_2)=\delta(\DD_2-123 \mu K^2)$ & 1.1\% \\
 ILC: \, \, $p(\DD_2)=\delta(\DD_2-195 \mu K^2)$   & 3.0\% \\
 Wd & 4.0\% \\
 Vdfs  & 3.1\% \\

\hline
  \end{tabular}
\end{center}

\end{table}

\section{The quadrupole and octopole alignment}

In this section we will evaluate the statistical significance of
various statistical measures which have been used to claim
the alignment between quadrupole and octopole.

Once again, we would like to stress the a-posteriori nature of the
estimators discussed below. In an idealized scientific experiment,
estimators and methods to be used to distinguish between various
theoretical models are known before the results and thus the inferred
constraints are objective in the sense that methods are ``blind'' with
respect to the data. In practice, however, progress is often made by
spotting patterns in the data which are not predicted by the
theory. One must be careful, however, when interpreting results which
are concerned with attempts to quantify the statistical significance
of the spotted regularities. The estimators and methods used in the
latter case are by construction biased towards showing a positive
detection. In the remaining part of this paper we simply wish to
address the sensitivity of various estimators to the uncertainties
caused by foreground subtraction and sky cuts. We do not try to asses
the question of meaning of probabilities associated with various methods.

\subsection{Angular momentum dispersion}

Authors of \cite{2004PhRvD..69d063516} have defined a unique axis for
each multipole motivated by quantum-mechanical considerations. For
each $\ell$ the axis $\mathbf{v}$ is defined as one that maximizes the
angular momentum dispersion (AMD) given by
\begin{equation}
  K = \sum_m m^2 |a_{\ell m}|^2.
\end{equation}

 By examining the absolute value of the dot-product between the AMD
vectors for quadrupole and octopole 
\begin{equation}
  d = | \mathbf{v_{\ell=2} \cdot v_{\ell=3}} |,
\end{equation}
 one can asses the statistical significance for their alignment.  The
absolute value takes into account the fact that these vectors are
head-less (i.e. their negative also maximize $K$). It can be easily
shown that the distribution for $d$ is uniform between zero and one
if AMD vectors are randomly distributed on the sky.

We have implemented the code that finds the AMD vectors for quadrupole
and octopole using the matrix transformations that can be found in the
appendix D of \cite{2004PhRvD..69d063516}. We determine the
maximum value of $K$ by numerical maximization rather than calculating
the values of $K$ on a Healpix grid. Using values for $a_{2m}$ and
$a_{3m}$ from \cite{2004PhRvD..69d063516} we are able to reproduce
their value of $d=0.986$.

\begin{figure}
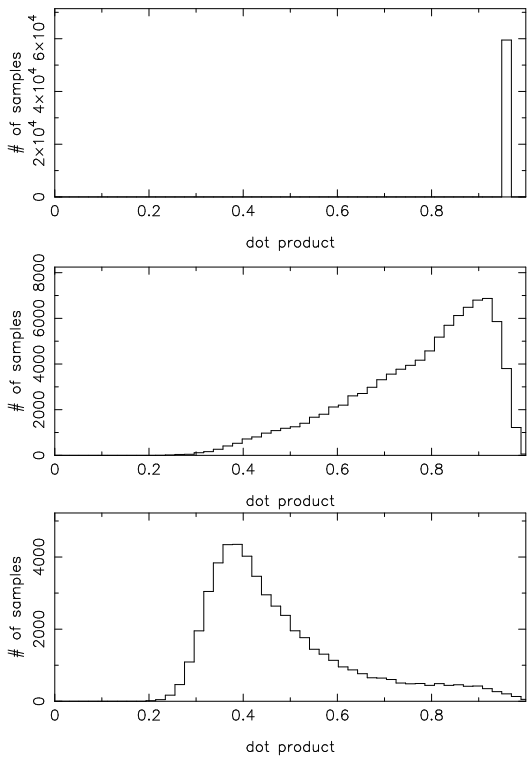

\epsfig{file=arf1.eps, height=0.8\linewidth, angle=-90}
\epsfig{file=arf2.eps, height=0.8\linewidth, angle=-90}
\epsfig{file=arf3.eps, height=0.8\linewidth, angle=-90}

\caption{\label{fig:teggy} The probability distribution function for
  the alignment of quadrupole and octopole using the maximum angular momentum
  dispersion method. The value of $1$ indicates fully aligned
  vectors, while a random model would correspond to uniform probability 
distribution. See text for further discussion.}
\end{figure}

When this method is applied to our MCMC chains we obtain probability
distribution function for $d$. This probability distribution function
can, in principle, be used to compare the Bayesian evidence for
aligned quadrupole and octopole model ($d=1$) with a standard model
($d$ being uniform between 0 and $1$). We plot our results in Figure
\ref{fig:teggy}. These results are worthy some discussion. First, in
the case of ILC map we get the value of $d=0.95$ which is consistent
with \cite{2004PhRvD..69d063516} and the small difference between
$\sim 0.95$ and $\sim 0.986$ is caused by the difference between the
WMAP ILC map and the full-sky TOH map. For the Wd case, we see that
the moderately high values of $d$ are still preferred, but the
distribution develops a large tail towards smaller values of $d$. It
is interesting, however, that the very high values of $d$ are strongly
disfavored. Values as high as $0.98$ are allowed, but the values of $d
\sim 1$ seem to be strongly rejected. In the Vdfs case all evidence
for high values of $d$ seem to vanish.  The probability for $d>0.98$
is 0.11\% for Wd and 0.5\% for Vdfs, compared to 2\% probability for
the random distribution.  The corresponding numbers for $d>0.95$ are
2.3\% and 1.6\%, compared to 5\% probability if random. The
probability for alignment exceeding these two values in the data is
below what it would be in the complete absence of the data (random
case). Thus the data do not show any evidence for alignment once the
foreground uncertainties are included in the analysis.

A more detailed inspection reveals that the quadrupole vectors
remained fairly well defined, while the vectors for octopole developed
a strong plane degeneracy.  This degeneracy is responsible for the
decrease in the statistical evidence for alignment. The main
conclusion from this investigation is that the alignment is not robust
against different treatments of foreground subtraction and that a
complete alignment is strongly disfavored by the data.

\subsection{Multipole Vectors}

Another method to explore the alignments are the multipole vectors
\cite{2003astro.ph.10511C}. It is based on the idea that every
multipole of the order $\ell$ is fully determined by $\ell$ headless
vectors $\mathbf{\hat{v}}^{\ell,i}$ such that

\begin{equation}
  \sum_{m} Y_{\ell m} \left(\hat{e}\right) a_{\ell m} = A^{(\ell)}
  \prod_{i=1}^{\ell} \left(
  \mathbf{\hat{v}}^{\ell , i} \cdot \mathbf{\hat{e}}\right).
\end{equation}

Pairs of these vectors can be used to form oriented areas, by taking a
cross-product between them:

\begin{equation}
  \mathbf{w}^{\ell, i,j} =  \mathbf{\hat{v}}^{\ell , i} \times  \mathbf{\hat{v}}^{\ell , j}.
\label{w}
\end{equation}

We thus have one such ``area'' vector for the quadrupole and three for the
octopole. If one takes the dot-products between $\mathbf{w}^{2,1,2}$
(quadrupole) and $\mathbf{w}^{3,i,j}$ (three octopole vectors) and
orders them in decreasing magnitude one obtains three numbers denoted
$A_1$, $A_2$ and $A_3$. A recent paper \citep{2004astro.ph..3353S}
claims that these values are anomalously high, indicating the
alignment between quadrupole and octopole. 
Since the multipole vectors are well defined only on full sky the 
analysis has been applied to full sky ILC maps, which may have residual 
contamination due to imperfect foreground subtraction. 
Our method is ideal to address these concerns in a systematic fashion. 

\begin{figure}
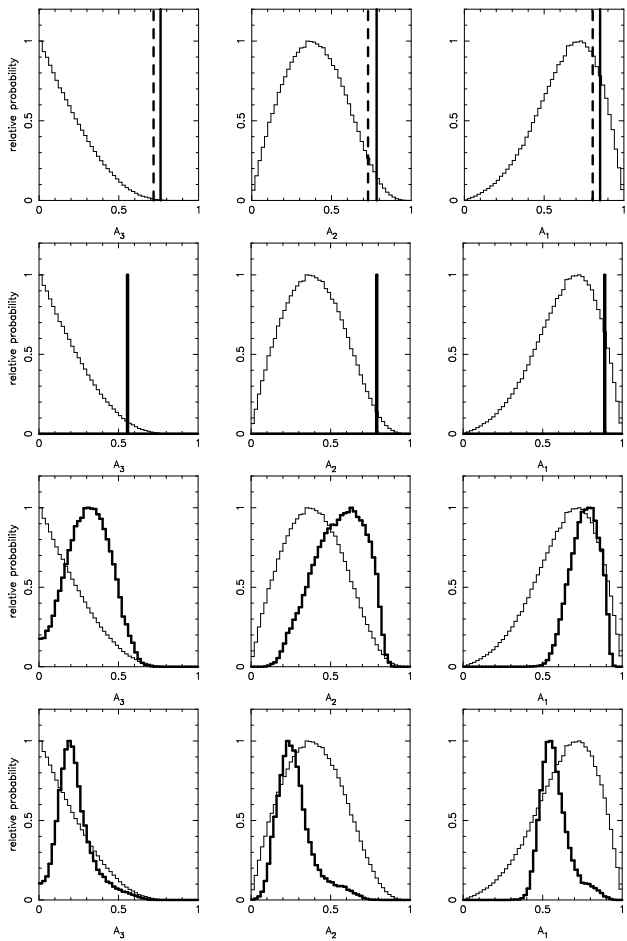

\epsfig{file=mpv2.eps, height=0.95\linewidth, angle=-90}
\epsfig{file=mpv11.eps, height=0.95\linewidth, angle=-90}
\epsfig{file=mpv12.eps, height=0.95\linewidth, angle=-90}
\epsfig{file=mpv13.eps, height=0.95\linewidth, angle=-90}

\caption{\label{fig:mpv} The probability distribution for $A_i$
  measures of alignment using multipole vectors. The top line shows
  the prior distribution for $A_3,A_2,A_1$ (from left to right) and
  the position of corresponding values for the dynamic quadrupole
  corrected (solid) and the dynamic quadrupole uncorrected (dashed)
  TOH maps \cite{2003PhRvD..68l3523T}. The next three rows correspond
  to WMAP ILC, Wd and Vdfs case respectively with thin line
  corresponding to the prior distribution and the thick line the
  distribution upon the addition of the data. The most likely point is
  normalized to one.}

\end{figure}

We have calculated the multipole vectors for our chains using the
publicly available code \citep{2004astro.ph..3353S}. We used these
vectors to obtain the probability distributions for $A_i$ in all cases
of our chains, as well as on a Monte-Carlo simulation of 150000 random
realizations of the sky to get the null-hypothesis distribution for
$A_i$. The results are plotted in Figure \ref{fig:mpv}.  Applying the
method to the full sky TOH map we are able to reproduce the results in
\cite{2004astro.ph..3353S}. Note that the results are more significant
for the dynamic quadrupole corrected than dynamic quadrupole
uncorrected map (see \cite{2003PhRvD..68l3523T} for detailed
description of the differences between these).  All
of the $A_i$ values, and $A_3$ in particular, are high.  Applying this
analysis to ILC map we find similar, although somewhat less
statistically anomalous, results.

We investigate next the effect of foreground uncertainty on these
statistics.  It is clear from figure \ref{fig:mpv} that introducing these
uncertainties significantly degrades the statistical significance.
%Note that the distributions do not extend into the anomalously
%high $A_i$ region. 
This is particularly clear for $A_3$, which has a value in excess of
0.7 for TOH, while neither Wd not Vdfs probability distributions
extend this high.  The probability for $A_3>0.7$ is 0.09\% for Wd and
0.02\% for Vdfs, compared to 0.2\% for the random case. Once again,
adding the observations {\it reduces} the probability of an alignment
(defined here as $A_3>0.7$) relative to no observations. This is not
very sensitive to the value we choose for alignment, i.e. we find
similar effect for $A_3>0.6$ and $A_3>0.5$.  Other parameters give similar
results, as is clear from Figure \ref{fig:mpv}.  We thus see no
evidence that these parameters are anomalously high in the data
compared to the random case, once the foreground uncertainties are
accounted for.  Thus, any evidence for the anomalously high value in
the full sky map must come from the region that is strongly affected
by the sky cuts or foreground subtraction. In other words, the
evidence for the anomaly comes from the region that is most likely to
be contaminated by foregrounds and so is not a robust feature in the
data.

%%%%%%%%%% matching features plot
\begin{figure}
\epsfig{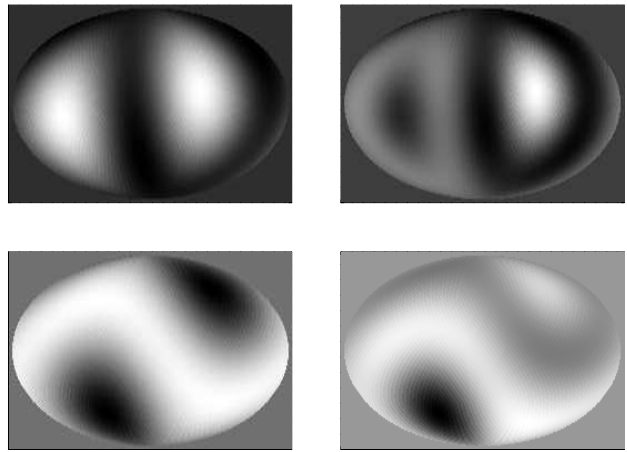}

\caption{\label{fig:primer} Two examples of a
  quadrupole (left) and octopole (right) realizations of a random sky
  that have particularly high value of the $I$ parameter.}

\end{figure}

\begin{figure}
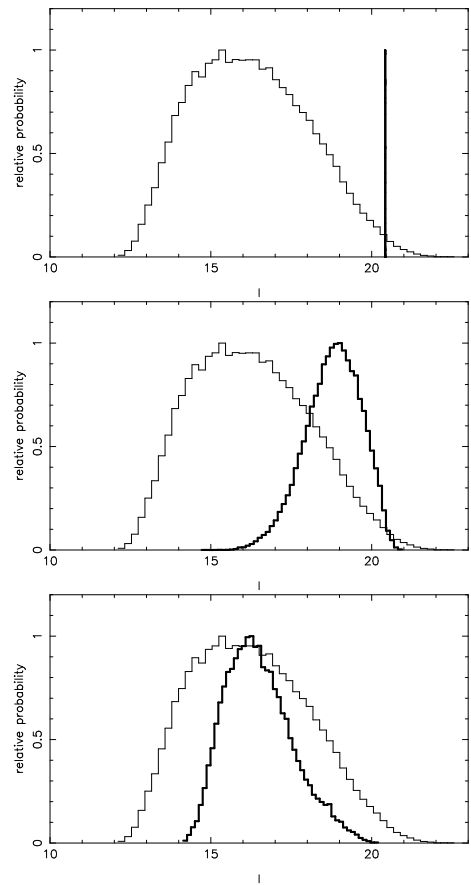

\epsfig{file=match1.eps, height=0.7\linewidth, angle=-90}
\epsfig{file=match2.eps, height=0.7\linewidth, angle=-90}
\epsfig{file=match3.eps, height=0.7\linewidth, angle=-90}

\caption{\label{fig:match} The probability distribution for the  $I$
 parameter for the ILC, Wd and Vdfs cases respectively. The thin
 line corresponds to the prior distribution, while the thick line to
 the distribution upon the addition of the data. The most likely point
 is normalized to one.}

\end{figure}

\subsection{Feature matching}

As a third example of alignment statistic we present a new method for
 determining alignment of the quadrupole and octopole. If the two are
 aligned, one expects that hot-spots and cold-spots match between the
 two. We thus examine the cross-correlation map created by multiplying
 the quadrupole and octopole maps:
\begin{equation}
T_{\times} = \hat{T}_{\ell=2} \times \hat{T}_{\ell=3}.
\end{equation}

These maps are created by first normalizing $a_{\ell m}$ so that
$\DD_2=1$ and $\DD_3=1$ (this is indicated by hats in the above
equation). We do this, because we want to decouple the multipole
amplitude from the alignment effect.  The integral of this map across
the sky necessarily gives zero due to the orthogonality condition:
\begin{equation}
  \int T_{\times} {\rm d}A = 0.
\end{equation}

However, the integral of the cross-correlation map gives a non-zero
result 
\begin{equation}
  I = \int \left( T_{\times}^2 \right) {\rm d}A,
\end{equation}
which is particularly high if the cold-spots and hot-spots match
between the two. This is illustrated in Figure \ref{fig:primer}
where we show two examples of a random sky realization with a
particularly high value of the $I$ parameter. The figure indicates
that the method indeed captures the heuristic idea of 
aligned quadrupole and octopole in terms of hot and cold spot overlap.

We calculate the value of $I$ parameter for all three
cases and also a large number (150000) of Monte Carlo simulations of a
random sky to determine the prior distribution of the $I$
parameter. The results are shown in Figure \ref{fig:match}.
The value for the $I$ parameter in the
case of the ILC map is again very high, with the 
random probability of being of that value or 
higher of only 0.9\%. Once again
this anomaly disappears in the case of the more conservative approaches, 
suggesting that the alignment is in the region most contaminated by the 
foregrounds.

\subsection{Alignment with the Ecliptic pole}

\begin{figure}
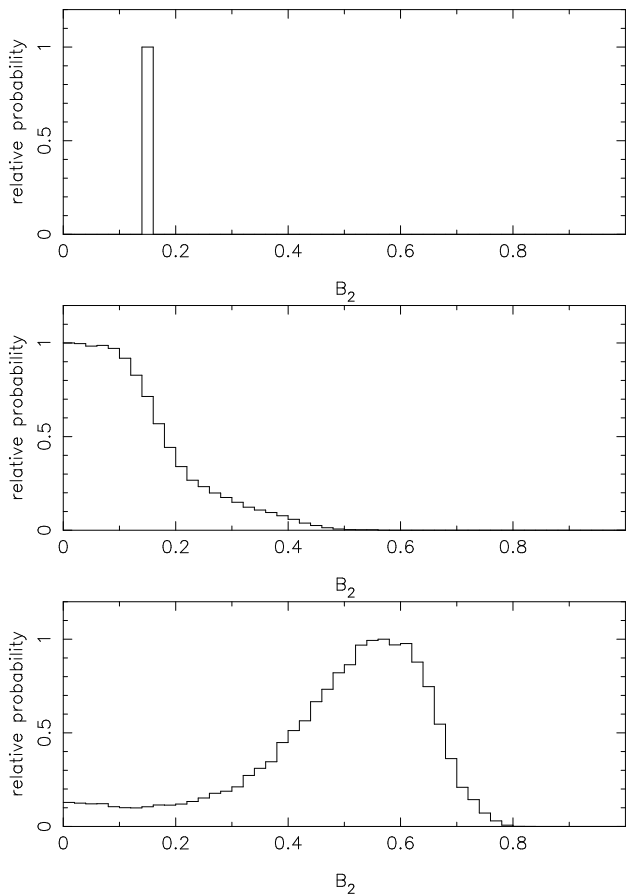

\epsfig{file=ecl21.eps, height=0.95\linewidth, angle=-90}
\epsfig{file=ecl22.eps, height=0.95\linewidth, angle=-90}
\epsfig{file=ecl23.eps, height=0.95\linewidth, angle=-90}

\caption{\label{fig:ecl1} The probability distribution for the $B_2$
  parameter. The prior distribution is flat. See Figure \ref{fig:ecl2}
  for the distribution of the $B_{3i}$ parameters. The panels
  correspond to ILC (top), Wd (middle) and Vdfs (bottom).}

\end{figure}

\begin{figure}
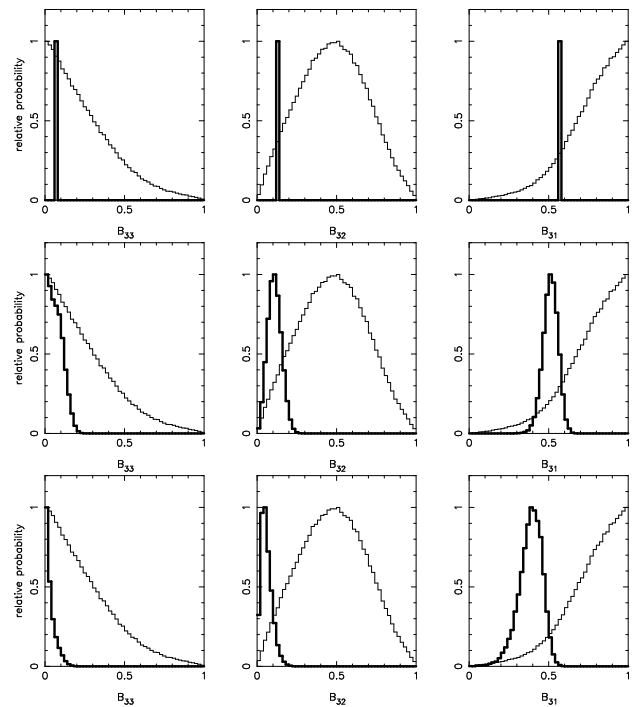

\epsfig{file=ecl11.eps, height=0.95\linewidth, angle=-90}
\epsfig{file=ecl12.eps, height=0.95\linewidth, angle=-90}
\epsfig{file=ecl13.eps, height=0.95\linewidth, angle=-90}

\caption{\label{fig:ecl2} The probability distribution functions for
  $B_{3i}$ (left to right) for the ILC (top), Wd (middle) and Vdfs
  (bottom) datasets.  The most likely point is normalized to one. }

\end{figure}

As a final example of the application of our method we investigate the
asymmetry in the WMAP data on two hemispheres separated by the
ecliptic plane. There were several reports of detection of this
asymmetry using various methods
\cite{2003astro.ph..7507E,2004astro.ph..2396H,2004astro.ph..4206H,2004astro.ph..3353S,2004astro.ph..4037L}.
The method most relevant for us is the angle between multipole
$\mathbf{w}$ vectors, defined in the Equation \ref{w}, and the
ecliptic north pole \cite{2004astro.ph..3353S}.  This test is
performed by calculating the dot product of the $\mathbf{w}^{\ell,
i,j}$ vectors with the north ecliptic pole. We perform this
multiplication for the quadrupole and call the resulting quantity
$B_2$. When the same is performed for the octopole, we sort the values
in ascending order and call them $B_{31}$, $B_{32}$ and $B_{33}$.
Results of these analyses for our MCMC chains are shown in Figures
\ref{fig:ecl1} and \ref{fig:ecl2}. 

%% As is clear from figure \ref{fig:ecl1} there is nothing anomalous in
%% quadrupole and there are very large differences between Wd and
%% Vdfs.

We see that these results are considerably more stable with respect to
the sky cuts and foreground marginalization than alignments
discussed above. In order to asses this, we introduce three
models:

\begin{itemize}
  \item \texttt{NULL} model assumes that $B$ parameters are
  distributed according to the random sky hypothesis
\item \texttt{ALIGN1} model assumes that $B_{31}<0.02$ and
  $B_{32}<0.02$. This is effectively equivalent to the assumption that
  $B_{31}=B_{32}=0$, but with the advantage that its Bayesian evidence can
  be calculated from the existing chains.
\item \texttt{ALIGN2} model assumes $B_{31}<0.02$,  $B_{32}<0.02$ and
  additionally $B_{2}<0.02$.

\end{itemize}

These models are, of course a posteriori. In fact, there are 16
possible models similar to \texttt{ALIGN1} and \texttt{ALIGN2} in
which one or more $B$ are anomalously low. Therefore, interpretation
of any result coming from the above should consider the the 1 in 16 factor
coming from the biased choice of models. If we take into account the
fact that one is working with the derived $\mathbf{w}$ vectors rather
than the source multipole vectors $\mathbf{v}$ the biased choice
``factor'' becomes even higher.

Nevertheless, we asses the probability of this occurring by chance
using two methods. Firstly, we measure the ratio of probabilities of
each model given data to its probability in the isotropic case and
secondly we calculate the Bayesian evidence for all three models for
the Wd and Vdfs case.  Bayesian evidence is the probability of a given
model in the Bayesian context, assuming all models to have the same
prior probability. It is the likelihood integrated over the prior
volume and thus disfavors models that have either low likelihood or
unnecessarily large number of parameters (leading to large volume of
parameter space having low likelihood). For further discussion see
\cite{mackay2002}. See Appendix \ref{sec:appA} for discussion of our
method for estimating evidence.  Results are shown in Table 3.  The
two methods are different in the way how the \texttt{NULL} model is
treated (i.e. the $E_{\rm NULL}$ is sensitive to how well the data
describe the isotropic model), but they nevertheless give similar
results.  While ILC does not show evidence for alignment as we defined
it, the evidence for the alignment of the quadrupole and octopole
planes with the ecliptic increases with the addition of the sky-cuts
and foregrounds. This can happen if, for example, the real signal is
contaminated by foregrounds, which destroy its evidence, but once the
contamination is removed the signal comes out again. In Vdfs ALIGN1
model is 42 times more likely than the random model (which corresponds
to $\sim 2.5 \sigma$).  As discussed above, however, once the biased
nature of the models is taken into account, the real statistical
significance is much smaller.  The main point of interest here is that
the alignment with ecliptic is less sensitive to foreground
contamination than other alignments discussed above.  We also
calculated the dot product of the ecliptic north pole with the Angular
Momentum Dispersion vectors, but the results did not show any evidence
for alignment.


\begin{table*}
\noindent
{\footnotesize Table 3: This table shows the two statistical tests
  used to asses the statistical evidence  for \texttt{ALIGN1} and
  \texttt{ALIGN2} models as discussed in the text.}
\begin{center}
\begin{tabular}{ccccc}
\hline
    Dataset & $E_{\rm ALIGN1}/E_{\rm NULL}$ &  $\frac{p(\mbox{\tiny{ALIGN1}}|{\rm data})}{p({\mbox{\tiny ALIGN1}}|{\rm isotropic})}$ & $E_{\rm ALIGN2}/E_{\rm NULL}$ &  $\frac{p(\mbox{\tiny{ALIGN2}}|{\rm data})}{p({\mbox{\tiny ALIGN2}}|{\rm isotropic})}$ \\
\hline 
 ILC   &  $<10^{-3}$ & & $<10^{-3}$ & \\
 Wd & $2.8 \pm 0.04$ & $4.3 \pm 0.08 $ & $19.3 \pm 1$ & $19.0 \pm 1 $ \\
 Vdfs  &  $41.7 \pm 0.2$ & $63.9 \pm 0.3 $ &  $16 \pm 1.2$ & $15.6 \pm 1$ \\
\hline
  \end{tabular}
\end{center}

\end{table*}

\section{Conclusions}

We developed a method to incorporate the uncertainties in the
foreground removal and/or sky cuts into the statistical analyses that
otherwise require a full sky data. Our method is based on deriving the
full multi-dimensional likelihood distribution of the multipole
moments given the data and the foreground uncertainties, noise or sky
cuts. We compute this distribution using the MCMC sampling of the
likelihood function, which is very fast, since we only need to invert
the covariance matrix once.  We recommend these methods are used to
assess the statistical significance of an effect found in the full sky
maps, to verify the sensitivity of the result to foreground
uncertainties.

As an application of the method we have performed a detailed analysis
of various quadrupole and octopole statistics in the WMAP data. For
our standard procedure we use three datasets, ranging from the most
aggressive (and probably contaminated) full sky ILC/TOH map to the V
channel with applied KP2 mask and foreground marginalization of all
three major contaminants, namely the Dust, Free-Free and Synchrotron
emission.  We used MCMC method to determine the probability
distribution for the realization of the 12-dimensional quadrupole and
octopole multipole moments.

We wish to emphasize again that probabilities of any a-posteriori
statistic are always subjective (and some are more subjective than others,
since they were more tailored for the data at hand). For this reason
we focus here on the relative change in the probabilities between the
least conservative ILC full sky map analysis and the most conservative
map with galactic sky cut and foreground marginalization. The
motivation is that if the probability changes strongly between these
cases then it is not robust and may be contaminated by imperfect
foreground removal in ILC/TOH map.  This is of course not the only
possibility: it is always possible that the data happen to contain
important statistical information in the region most contaminated by
the foregrounds. But our method provides some objective estimate of
the probability for this to happen.

A point related to this is the question of how to formulate a
statistically meaningful quantity that addresses various claims in the
literature.  For example, one could test perfect alignment between
quadrupole and octupole ($d=1$), but a somewhat less perfect alignment
could also be of potential significance as a sign of an anomaly.  We
address this by computing the integrated probability for a statistic
to exceed a certain value and compare it to the random case. As the
chosen value approaches the limit (e.g. $d=1$) the probability for the
random case becomes very small, while if the data show signs of
anomaly the integrated probability from the real data will remain
significant. The ratio of the two thus gives some information on
whether there is anything anomalous in the data.

We analyze the evidence for the low quadrupole and octopole.  These
are not particularly unlikely from ILC map, with the quadrupole
probability being 3\% and octopole being quite typical.  These numbers
do not change much by our more conservative treatment, so they appear
robust. They are however significantly higher than the original values
quoted by WMAP \cite{2003astro.ph..2209S}, which are statistically
excluded and are therefore not a result of differences in foreground
modeling.  In a certain sense we have reached the opposite conclusion:
from our analysis it appears impossible for the octopole to be
anomalously low and the same is true for quadrupole in our Wd
analysis.  The difference may be a consequence of the noisier
estimator used by WMAP.

We also discuss recent claims that the quadrupole and octopole are
aligned. If one believes the ILC map, then the evidence for the
quadrupole and octopole alignment is considerable. All three methods
tested here, namely the maximum angular dispersion vectors, the
multipole vectors and the feature matching method indicate that the
two are suspiciously aligned. However, as soon as foreground
uncertainties are included the evidence for this alignment
disappears. It is not unexpected that the probability distributions
broaden, but what is surprising is how rapidly the evidence vanishes
and how strongly perfect or even partial alignment is excluded by the
data. This strongly suggests that much of the evidence of the
alignment comes from the portion of the data most contaminated by the
galactic foregrounds.

Plots shown in this paper indicate a considerable difference between
Wd and Vdfs cases. It is interesting to explore whether this
difference comes from the number of templates being marginalised over
or whether they are due to different frequency channels being
employed. To investigate this we repeat the analysis using W channel
data and marginalise over all three templates. The results are very
similar to the Wd case, indicating that the main difference is due to
different channel being employed.  This suggests there might be
additional systematic effects that are not handled properly even by
our more conservative treatment.  One possibility is that either V or
W channel MEM derived foreground maps are contaminated on the largest
scales, or that there are systematic contaminations in CMB maps
\cite{2004astro.ph..3353S}.  More work is needed
to explore these various possibilities.

Finally, we also present an example where our method can enhance the
statistical significance by removing the contamination which would
otherwise mask the evidence.  We show that for the alignment of
multipole vectors with ecliptic plane the statistical significance is
not lowered by the foreground uncertainties.  Once again, the
statistical significance of this effect is unclear, but at least it
seems clear that it is not significantly affected by the foregrounds
and may even be enhanced, once the foreground contamination is removed
from the data.

Our method is statistical and relies on the foreground templates to be
faithful representation of all components that contaminate the
data. Systematic uncertainties in the foregrounds translate in
systematic uncertainties in derived quantities. Although our method
provides a statistical framework for assessing the statistical
significance of various effects, the real improvement will come from
better understanding and modelling of the foregrounds. This goal can
be achieved using multi-frequency data combined with a better
understanding of physical processes involved. In this case a less
conservative treatment of the foregrounds may be possible. We have
tried some of these examples in our tests. Analyzing the original V or
W maps without sky cuts, but with foreground template marginalization,
causes MCMC sampler not to converge, which is indicative of a complex
likelihood in this case.  Similarly, using sky cuts but with no
foreground removal shows clear evidence of contamination and causes
the quadrupole and octopole to increase significantly
\cite{2004astro.ph..3073S}.  Subtracting the WMAP recommended
foregrounds and using $\lambda=1$ in equation \ref{eq:lambda} gives
results very similar to $\lambda=\infty$. Thus we believe our results
reflect the current uncertainties in the foreground subtraction and
suggest these may be responsible for many of the anomalies seen in the
WMAP data on large scales.

\section*{ACKNOWLEDGMENTS}
We acknowledge valuable discussions with the authors of
\cite{2004astro.ph..3353S} and C. Hirata for useful comments. We acknowledge
the use of the Legacy Archive for Microwave Background Data Analysis
(LAMBDA). Support for LAMBDA is provided by the NASA Office of Space
Science.  US is supported by Packard Foundation, Sloan Foundation,
NASA NAG5-1993 and NSF CAREER-0132953.

\appendix
\section{Evidence calculation}
\label{sec:appA}

The Bayes equation for the posterior probability of set of parameters
$\boldsymbol{\theta}$ given a data-vector $\mathbf{d}$ can be written as

\begin{equation}
  P(\boldsymbol{\theta}|\mathbf{d}) =
  \frac{P(\mathbf{d}|\boldsymbol{\theta}) P(\boldsymbol{\theta})}{P(\mathbf{d})}
\label{eq:3}
\end{equation}

The denominator of the right hand side is the evidence. See \cite{mackay2002} for
a discussion of the usage of evidence as a Bayesian tool for model
selection. Since the posterior must integrate to unity, one can write
\begin{equation}
\label{eq:ress}
  E = P(\mathbf{d}) = \int_{\boldsymbol{\theta}} P(\mathbf{d}|\boldsymbol{\theta}) P(\boldsymbol{\theta}),
\end{equation}
i.e. the evidence is the likelihood integrated over all
priors. Evidence for completely disjoint models is usually calculated
by the thermodynamic integration during the burn-in phase of the MCMC
sampling (see e.g. \cite{2003MNRAS.338..765H}). When considered models
are a subset of the most general model, it is possible to calculate the
relative evidence from the MCMC samples of the most general model.  We
perform the integral of the Equation (\ref{eq:ress}) in bins that are
$0.02$ in size over $B_2$, $B_{31}$, $B_{32}$. The integral is thus
performed by adding up the prior probability corresponding to each
sample. The prior probability is approximated to be constant over the
entire bin.  For the \texttt{NULL} model, the prior probability is
obtained by binning the Monte Carlo simulations of random skies on the
grid and normalizing. For the \texttt{ALIGN1} and \texttt{ALIGN2}
models, the prior probability is one in the corresponding bin
(i.e. $B_{31}<0.02$ and $B_{32}<0.02$ and zero otherwise). The prior
probability for $B_2$ is constant at $0.02$ for models \texttt{ALIGN1}
and \texttt{NULL}. For the \texttt{ALIGN2} it is one in the first bin
and zero otherwise.

The error on the evidence are assumed to be only due to Poisson error
in the number of samples in a given bin. If chains are well converged,
this should indeed be the dominant error.

\bibliography{../BibTeX/cosmo,../BibTeX/cosmo_preprints,mcmcbib}

\end{document}